\documentclass{elsart}
\usepackage{epsfig} 
\usepackage{subfigure}
\usepackage{amsmath}
\begin{document}
\begin{frontmatter}
\title{Characterization of volume type ion source for $p$, $H_2^+$ and $H_3^+$ beams}
\author[IAP]{N. Joshi\corauthref{cor}},
\corauth[cor]{Corresponding author.}
\ead{joshi@iap.uni-frankfurt.de}
\author[IAP]{M. Droba},
\author[IAP]{O. Meusel},
\author[IAP]{U. Ratzinger}.
\address[IAP]{Institut f\"ur Angewandte Physik, Goethe Universit\"at, Frankfurt am Main, Germany.}
\begin{abstract}
Recently, there is an increasing need for $H_{2}^+$ and $H_{3}^+$ ion sources. One  example are ion therapy facilities, where $C^{4+}$ and $H_{3}^+$ ion beams along the linac are of great interest. Another example is a $H_{2}^+$ test beam for linacs finally operated with intense deuteron beams. At Frankfurt, a simple proton ion source is needed to test a new kind of beam injection into a magnetic storage ring\cite{EPAC08}\cite{EPAC06}. This article describes a volume type ion source which can deliver upto $3.05~mA$ beam  current at $10~keV$ in stable dc operation. It is a hot filament driven ion source which can provide high fractions of $p$, $H_{2}^+$ or $H_{3}^+$, depending on the operation settings.
\end{abstract}
\end{frontmatter}
\section{Introduction}
Depending on the operating parameters, fractions of $p$, $H_{2}^+$ and $H_{3}^+$ are generated in the hydrogen gas plasma. $H_{2}^+$ and $H_{3}^+$ beams are used at fixed velocity heavy ion linacs like Unilac to get more beam intensity at a given space charge limit ($I_e=A/q$). At an energy above a few MeV the molecular ion beam can be converted into a proton beam at very high efficiency and with negligible loss in beam quality by passage through a thin Carbon stripping foil. Actually, $H_{2}^+$ and $H_{3}^+$ beams can be produced very efficiently by hot cathode driven volume plasma sources at modest operation parameters \cite{Volk}.

Applications might be at medical ion beam facilities for cancer treatment. If $Carbon$ and $p$ beams are needed, both beams are generated in identical Electron Cyclotron Resonance ECR-sources \cite{Maier}. The sources and injector settings have to be changed due to the $A/q-$ difference between $p$ and $C^{4+}$. Moreover, this type of ECR source was optimized for heavy ion beams. This paper explains, why it might be reasonable to add a third source to the injector complex, which can produce $H_3^{+}$ efficiently. The source is cheap , compact, and easy in maintainance. It allows a factor $3$ higher $p$ beam current at injector parameters identical to $C^{4+}$ operation down to the stripper foil - due to the space charge limit of a multicell linac.

Another application are high current $d$ and $t$ beam facilities \cite{Garin}. As these beams show a high activation potential it is very important to have a "start up " beam for finding operation parameter settings. This could be done very nicely with intense  $H_2^{+}$ and $H_3^{+}$ beam, respectively.

\section{Ion Source and Experimental Setup}
%=== Figure =======================
\begin{figure}[!h]
\vspace{0.5cm}
\begin{center}
\includegraphics*[width=12cm]{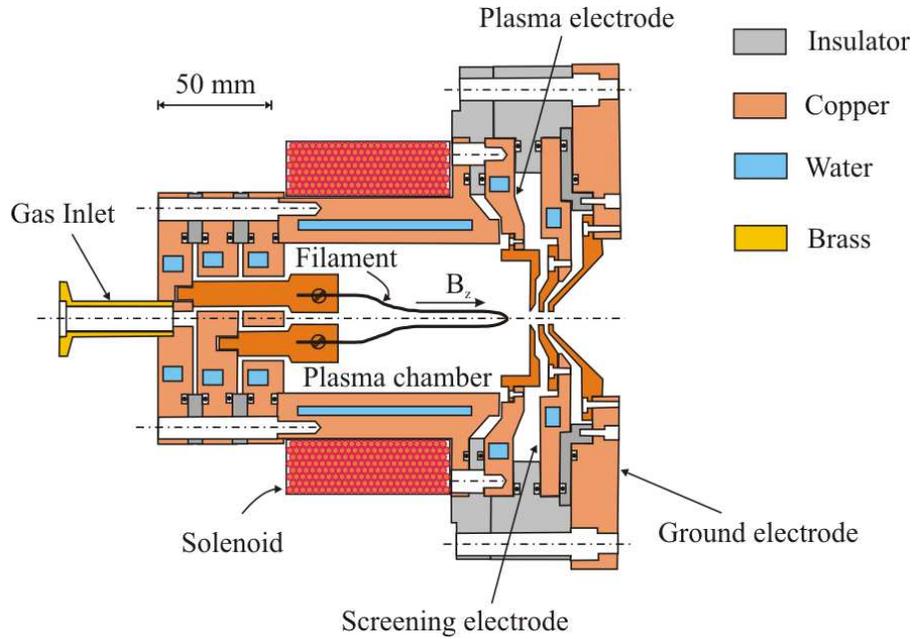}
\end{center}
\caption{Cross sectional view of ion source.}
\label{source_cross}
\end{figure}
\begin{figure}[hhh]
\vspace{0.5cm}
\begin{center}
\includegraphics*[width=14cm]{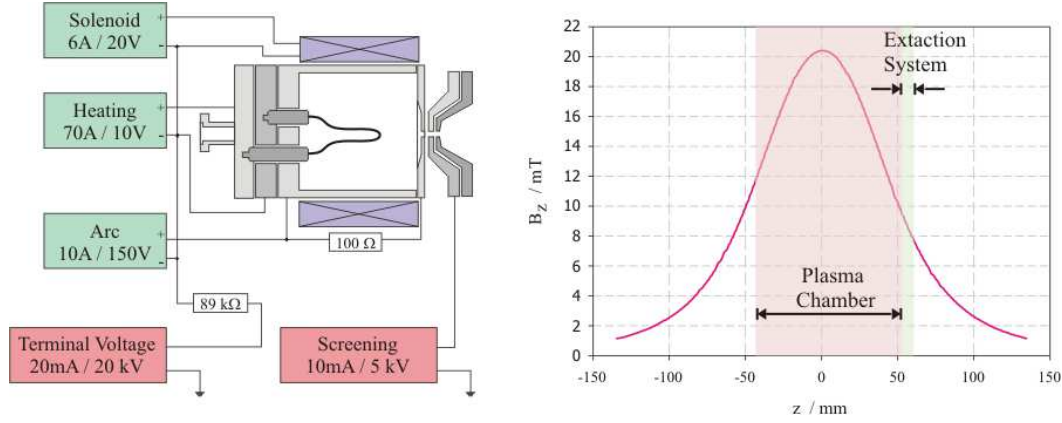}
\end{center}
\caption{Block diagram of source-circuit (left) and  magnetic field distribution of solenoidal coil (right).}
\label{source_mag}
\end{figure}

%=== Figure =======================
As shown in fig.\ref{source_cross} the source consists of a plasma chamber with a hot tungsten filament and a triode extraction system. The cathode emits electrons which are accelerated to the cylindrical plasma chamber wall. Electrons are forced into cycloidal paths by an external solenoidal field, increasing their path lengths  to the wall, and thereby, increasing the probability of an ionizing collision with neutrals. Fig.\ref{source_mag} shows the block diagram of the source with power supplies and the on axis magnetic field produced by the solenoid. One can see, that almost a $40\%$ magnetic field level is present at the extraction system. Magnetic field levels at the plasma chamber centre up to $23~mT$ were applied by coil currents up to $6~A$.  The heating filament as well produces a comparable magnetic field of around $9~mT$  by heating current upto $60~A$. This magnetic force does not have a significant effect on the ion beam, but does change the plasma properties significantly.
%=== Figure =======================
%=== Figure =======================
A triode extraction system was used. At normal settings the plasma electrode is held at positive potential, the screening electrode is held at a negative $10\%$ level of the plasma electrode potential. The space charge dominated beam current is given by the Child Langmuir Law,
% === Equation =================================================
	\begin{equation}
	J=\frac{1}{9\pi}~ \sqrt{\frac{2e}{m}}~\frac{V^{3/2}}{d^2}. 
	\label{KV}
	\end{equation}
%===  Equation  ================================================
$J$ is the current density, $V$ is the accelerating potential and $d$ is the acceleration gap length. In the source used $d$ was $5~mm$.
%=== Figure =======================
\begin{figure}[hhh]
\vspace{0.5cm}
\begin{center}
\includegraphics*[width=12cm]{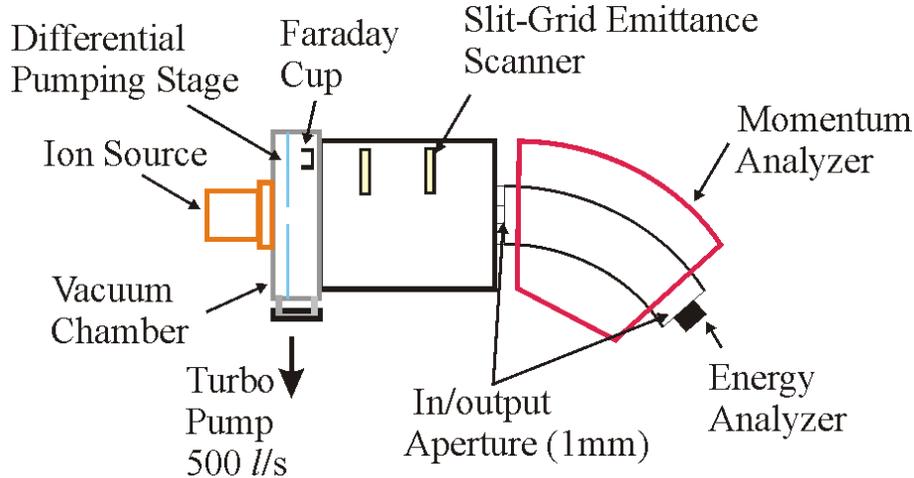}
\end{center}
\caption{Top view of the experimental setup.}
\label{setup}
\end{figure}
%=== Figu^re =======================
Fig.\ref{setup} shows the experimental setup used for a characterization of the ion source. The source was mounted on a differential pumping chamber equipped with three turbo molecular pumps. A Faraday cup was installed in the same tank for beam current measurements. An emittance scanner of the grid slit type was mounted and  a momentum analyzer was installed downstream.

\section{$He^+$ beam}
The helium beam consists of a single specie, singly charged ion beam for this source type, generated by the reaction 
% === Equation =================================================
	\begin{equation}
	He~+~e^-\longrightarrow~ He^+  ~+ 2e^- 
	\label{He_reaction}
	\end{equation}
%===  Equation  ================================================
% === Figure =================================================
\begin{figure}[!h]
\vspace{0.5cm}
\begin{center}
\subfigure[]{
\includegraphics*[width=4.2cm]{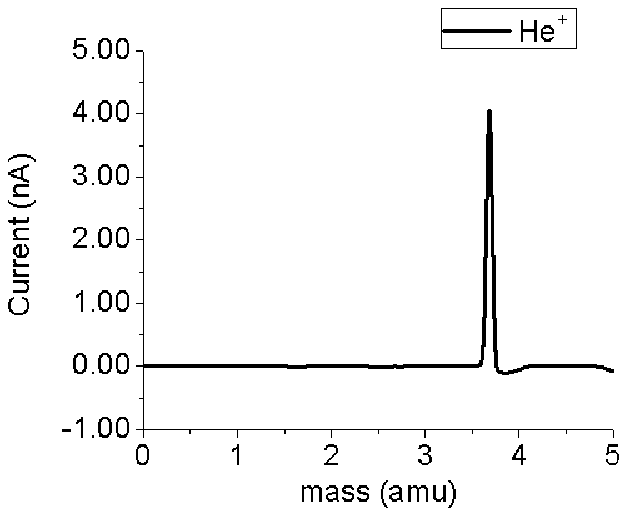}
}
\subfigure[]{
\includegraphics*[width=4.2cm]{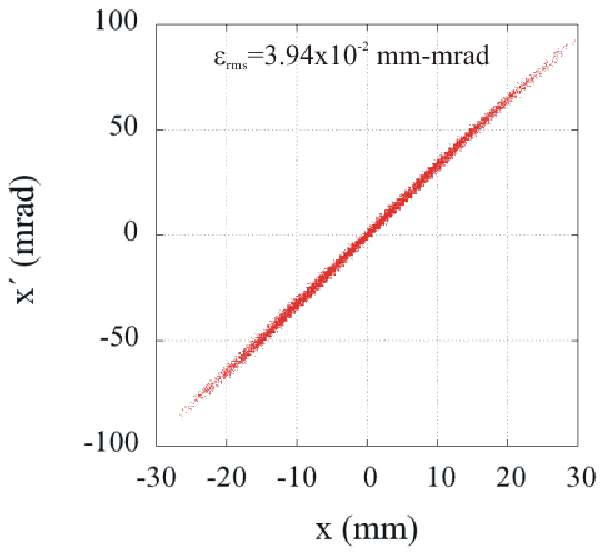}
}
\subfigure[]{
\includegraphics*[width=4.2cm]{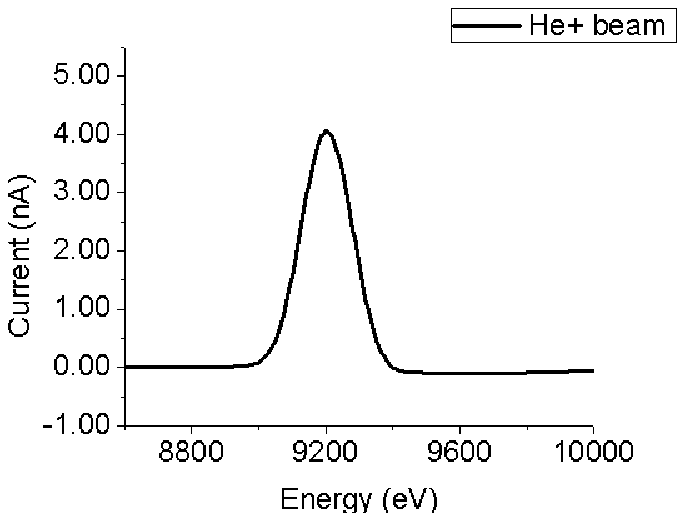}
}
\end{center}
\caption{(a) Graph showing the single specie of mass $m=4~amu$, (b) the phase-space distribution measured using the emittance scanner, and (c) Energy Spectra with an energy spread of $120~eV$ corresponding to $\pm1.5\%$ energy spread. The displacement in mass scale is explained as $m=1 amu$ was calibrated for $10~keV$ energy.}
\label{He_example}
\end{figure}
% === Figure =================================================
Fig.\ref{He_example} shows an example of a $He^+$ beam with current $1.2~mA$ extracted at $9.2~keV$ beam energy.  A gas filling pressure of $12.0hPa$ and $100~V$ arc potential were used as standard settings. The terminal can supply voltage levels up to $20~kV$.  The current extracted was measured by a Faraday cup.

\section{$p$, $H_2^+$, $H_3^+$ beams}
\begin{figure}[hhh]
\vspace{0.5cm}
\begin{center}
\subfigure[]{
\includegraphics*[width=4.2cm]{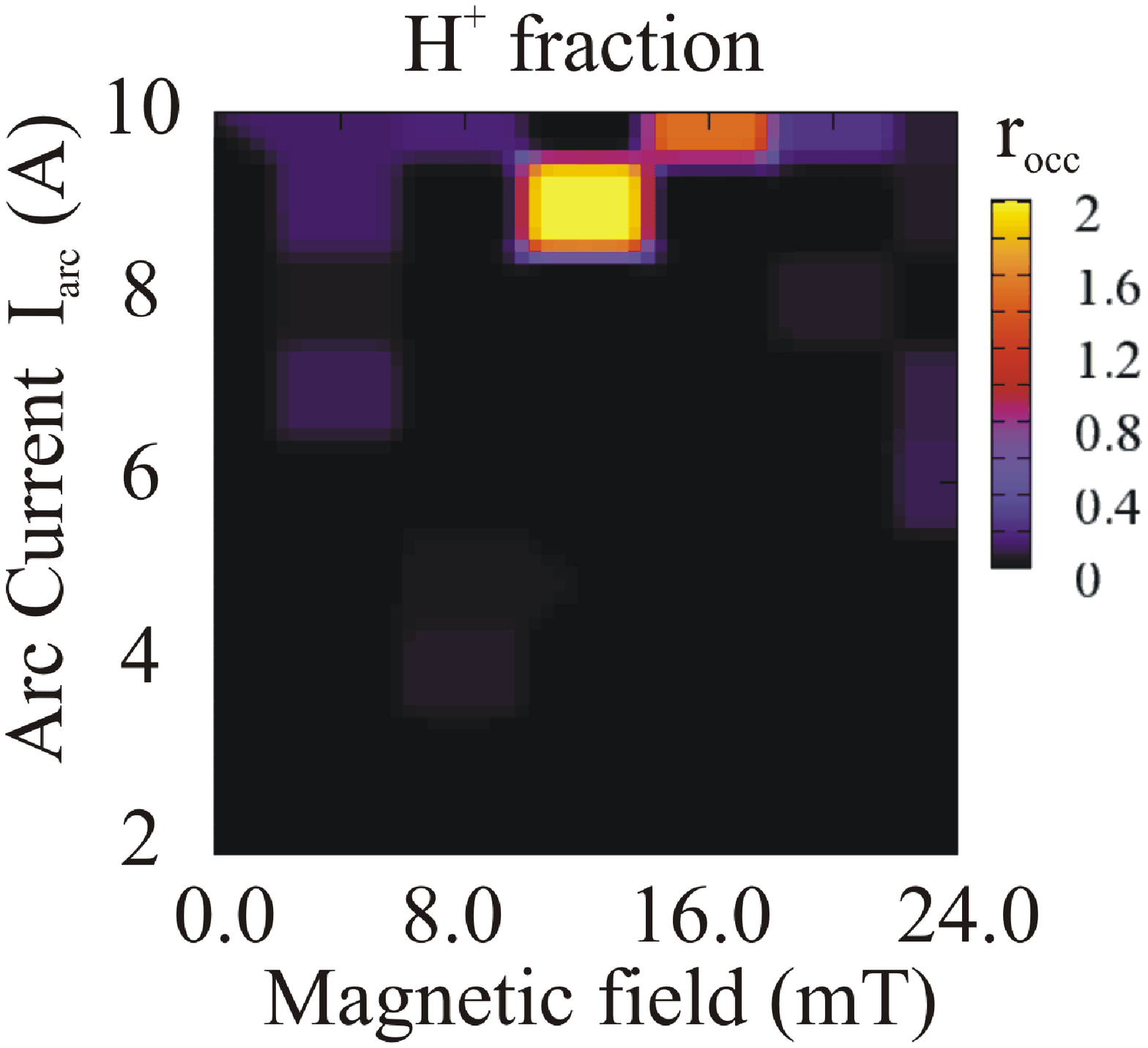}
}
\subfigure[]{
\includegraphics*[width=4.2cm]{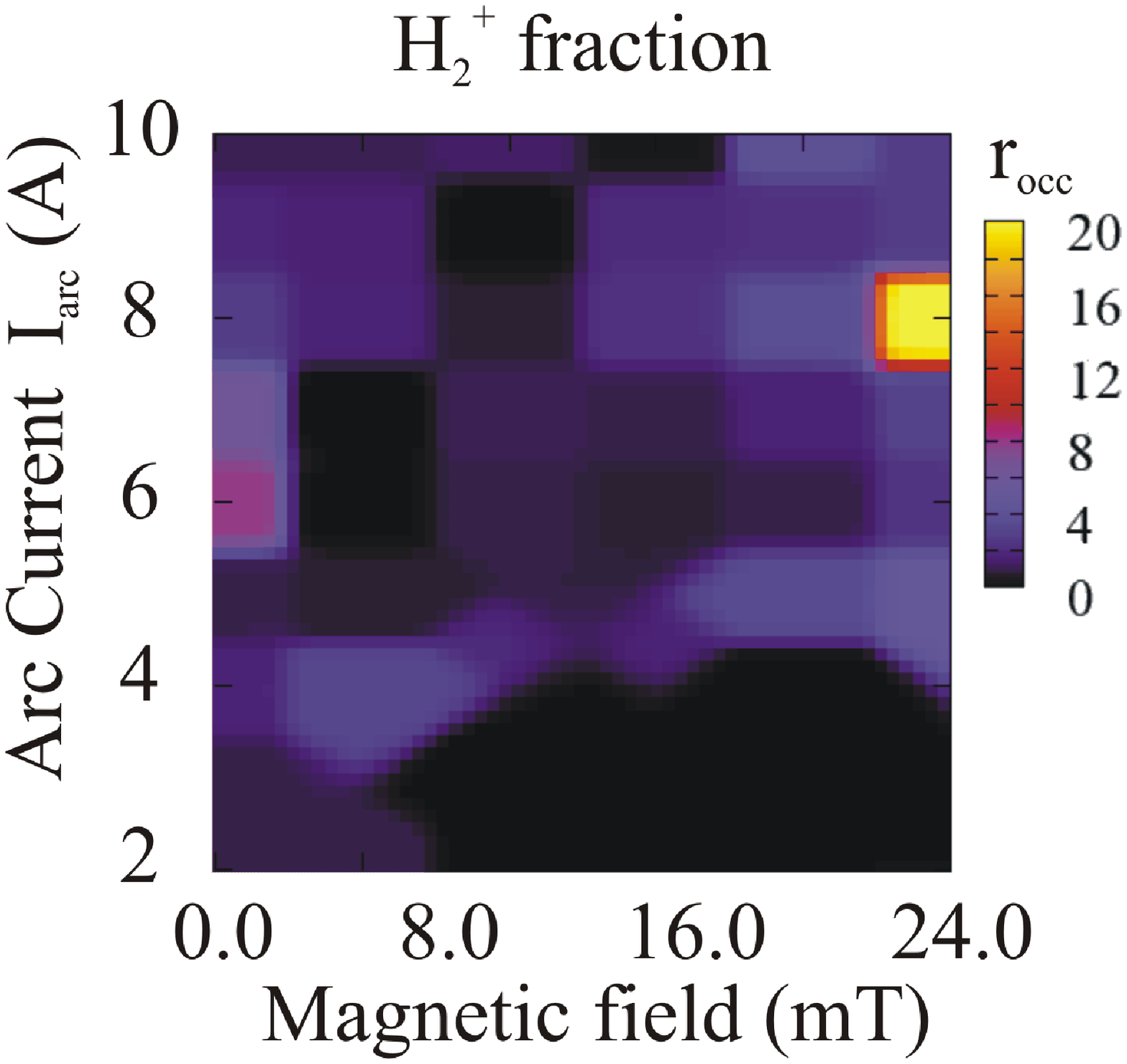}
}
\subfigure[]{
\includegraphics*[width=4.2cm]{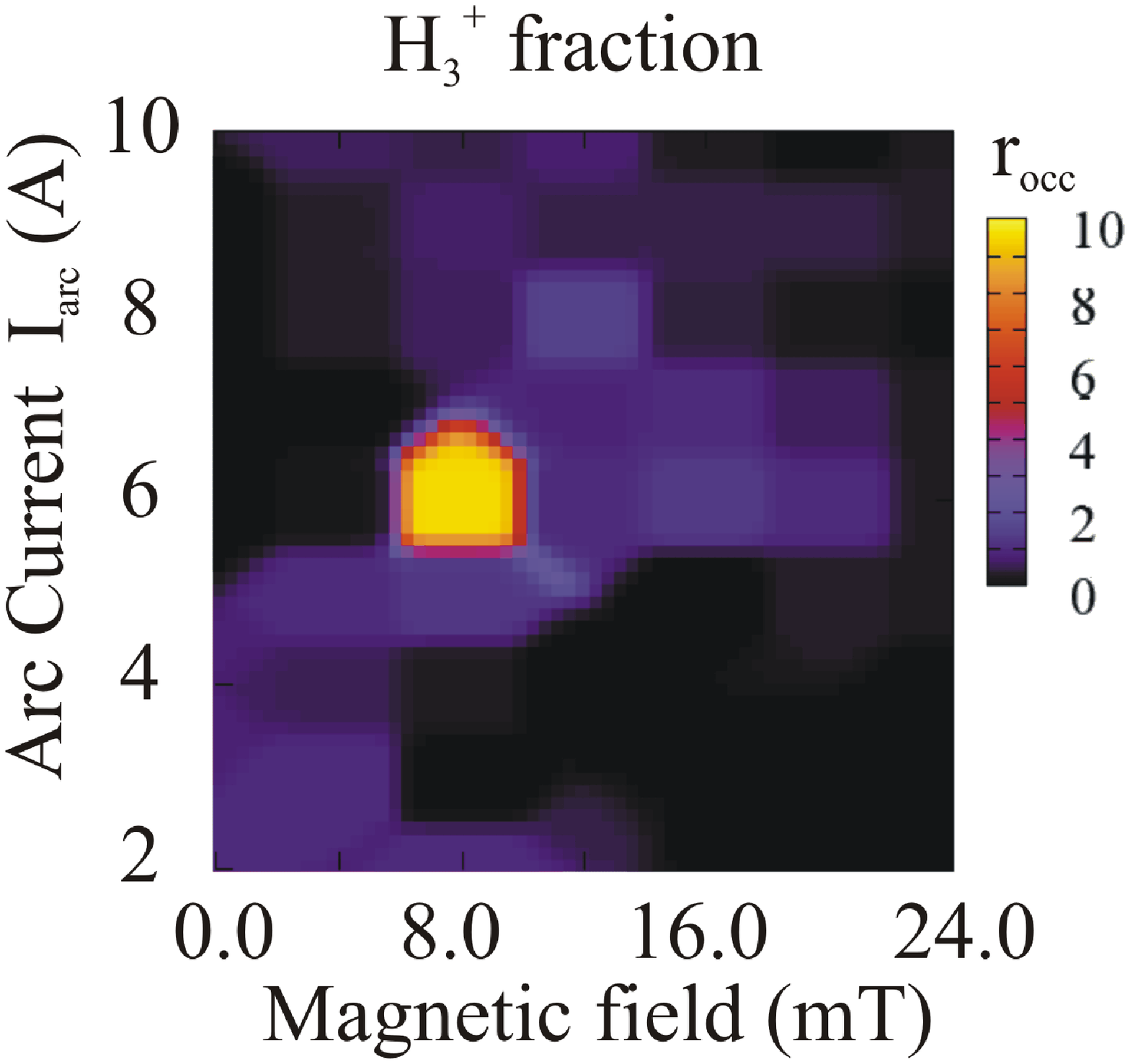}
}
\end{center}
\caption{Graph of relative occurrences of (a) $H^+$ , (b) $H_{2}^{+}$ and (c) $H_{3}^{+}$ plotted as a function of arc current ($I_{arc}$) and the maximum on axis magnetic field.}
\label{Rel_occ_max}
\end{figure}

The proton beam is extracted from a plasma fed by $H_2$. The key reactions involved are,
%== Equation =================================================
\begin{eqnarray}
~~~~~~~~~~~~~~~~~~~~~~	H_2 + e \Rightarrow H_2^{+} + 2e\nonumber\\
~~~~~~~~~~~~~~~~~~~~~~	H_2^{+} + e \Rightarrow H^{+} + H + e \nonumber\\
~~~~~~~~~~~~~~~~~~~~~~	H + e \Rightarrow H^{+} + 2e \nonumber\\
~~~~~~~~~~~~~~~~~~~~~~	H_2^{+} + H_2 \Rightarrow H_3^{+} + H \nonumber\\
~~~~~~~~~~~~~~~~~~~~~~	H_3^{+} + e \Rightarrow H^{+}+ H_2 + e.
\end{eqnarray}
%== Equation =================================================
The last two processes are assumed to be important for an efficient production of protons in a small source\cite{Groß}\cite{Holinger}\cite{Morishita}. It was figured out that the proton fraction depends upon plasma parameters. The whole parameter space of arc current and magnetic field was scanned at three different values of arc potential ($V_{arc}$) and three different values of source filling pressure ($P$) to find the maximum fraction of individual species. To get a clear contrast in the plots for identifying the optimum settings for the production of the species of interest, for example for protons, the \textit{relative occurrence} ($r_{occ}$) was defined as
% === Equation ==========================================
	\begin{equation}
	r_{occ}~(H^+)=\frac{\eta_{H^+}}{\eta_{H_{2}^{+}}+\eta_{H_{3}^{+}}},
	\label{rel_occ}
	\end{equation}
% === Equation =============================
where $\eta_{H^+}$ is the relative percentage at that perticular set of parameters. In a similar way, the $r_{occ}$ for $H_{2}^{+}$ and $H_{3}^{+}$ were defined. Fig.\ref{Rel_occ_max} shows the relative occurrence of $H^+$, $H_{2}^{+}$ and $H_{3}^{+}$ as function of arc current ($I_{arc}$) and magnetic field at the gas  filling pressure of $12~hPa$ and $80~V$ arc potential. At particular values of magnetic field and plasma density the relative occurrence of particular species is very high. Thus the graph shows an \textit{island} kind structure.
% === Figure =================================================
% === Figure =================================================
% === Figure =================================================
\begin{figure}[hhh]
\vspace{0.5cm}
\begin{center}
\subfigure[]{
\includegraphics*[width=4.2cm]{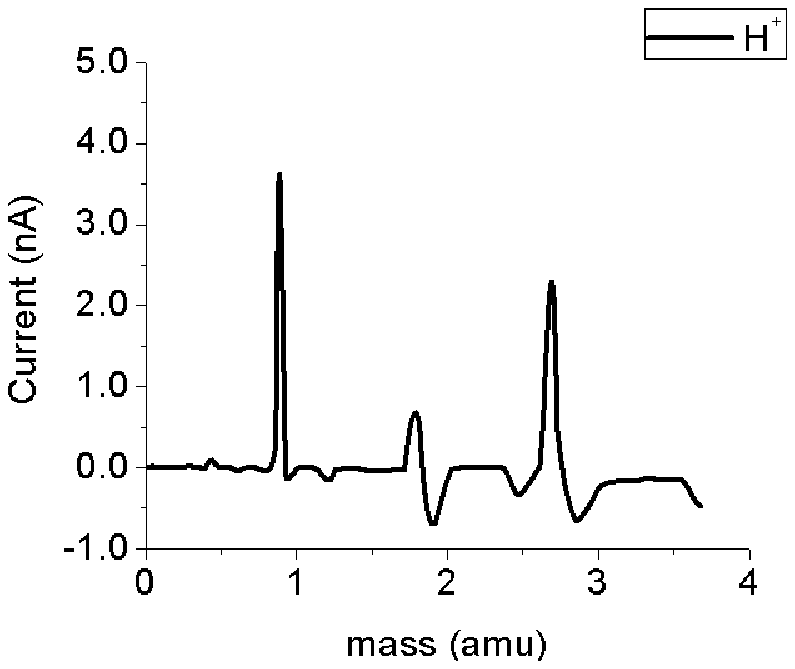}
}
\subfigure[]{
\includegraphics*[width=4.2cm]{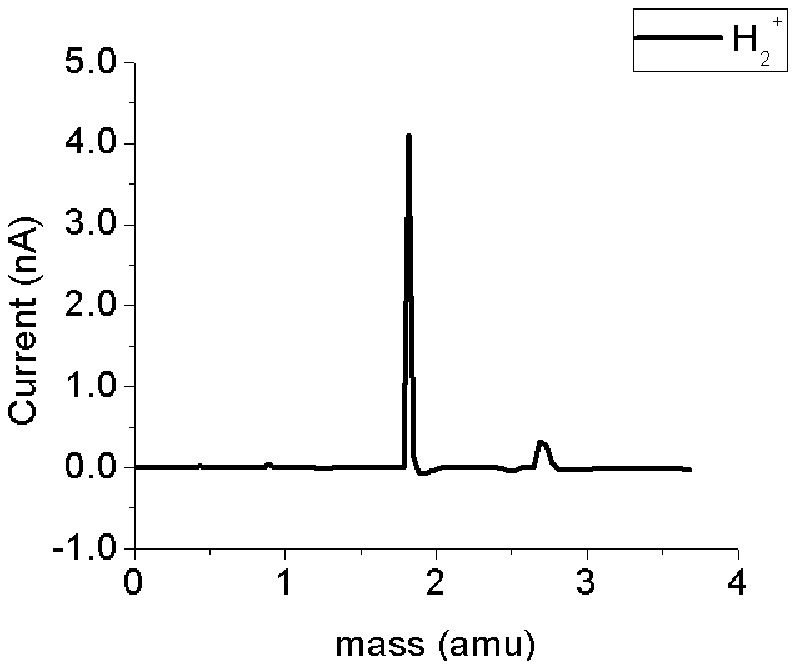}
}
\subfigure[]{
\includegraphics*[width=4.2cm]{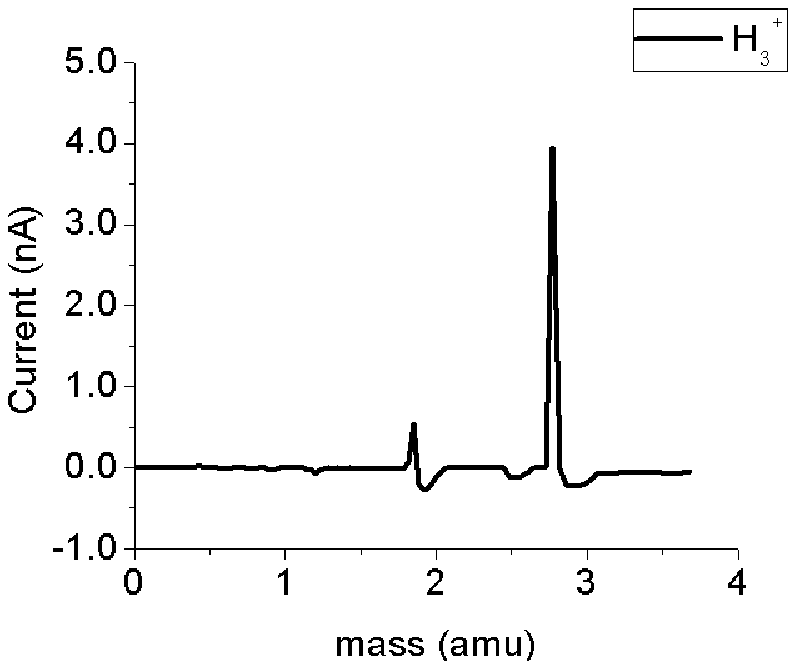}
}
\end{center}
\caption{Momentum spectometer graph showing the maximum peak for (a) $H^+$ , (b) $H_{2}^{+}$ and (c) $H_{3}^{+}$ for corresponding values of  arc current and magnetic field.}
\label{mass_spectra_max}
\end{figure}
% === Figure =================================================

Fig.\ref{mass_spectra_max} shows the mass spectra at the optimum values of relative occurrences. As optimum ratios at each of the separate \textit{island} positions, about $58\%$ of proton corresponding to $3.04~mA$, about $91\%$ of $H_{2}^{+}$ corresponding to $2.84~mA$ and about $95\%$ of $H_{3}^{+}$ corresponding to $3.05~mA$ current were extracted at $10~keV$ beam energy respectively.

\begin{figure}[hhh]
\vspace{0.5cm}
\begin{center}
\subfigure[$P=12.0~hPa$]{
\includegraphics*[width=4.2cm]{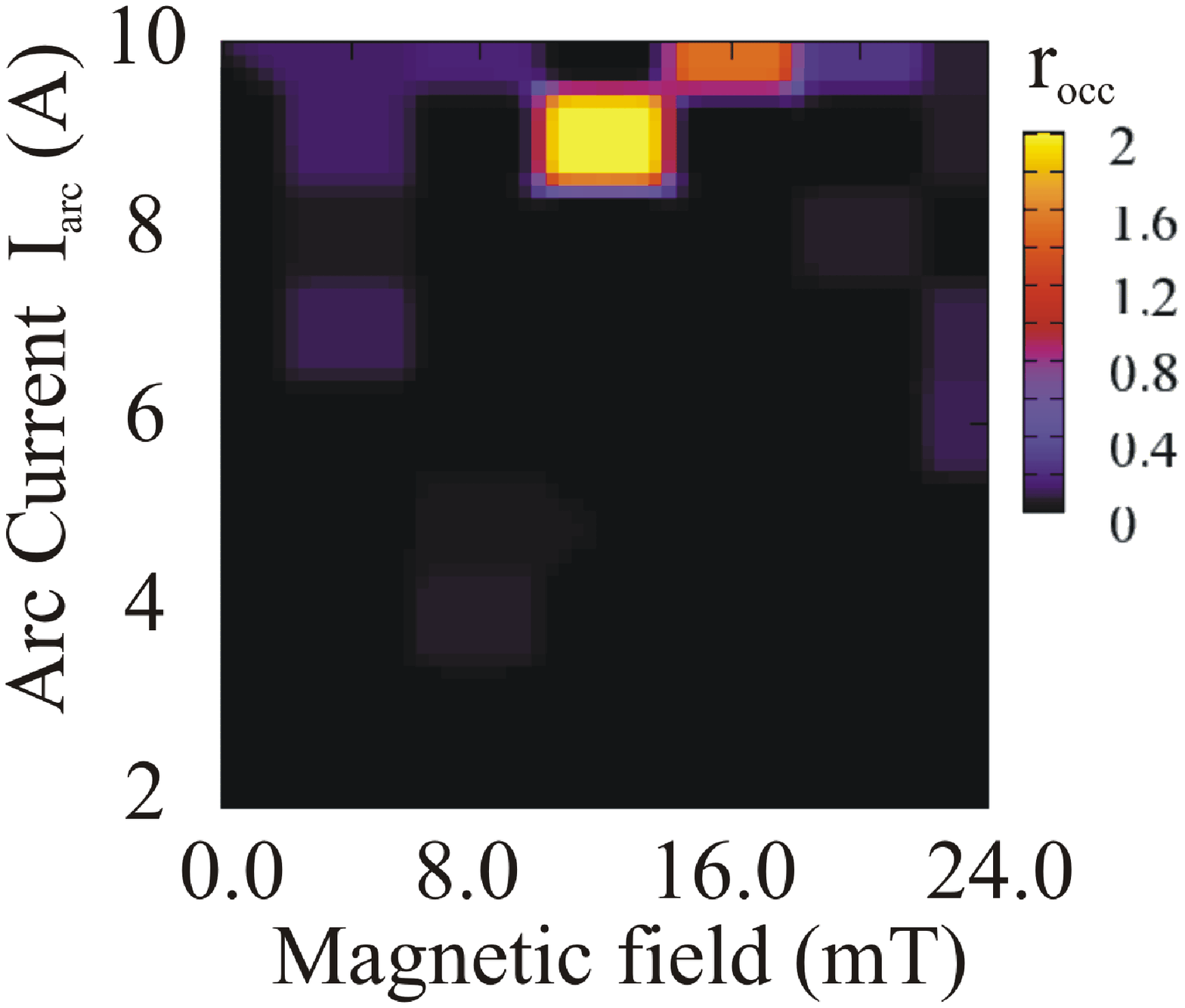}
}
\subfigure[$P=32.0~hPa$]{
\includegraphics*[width=4.2cm]{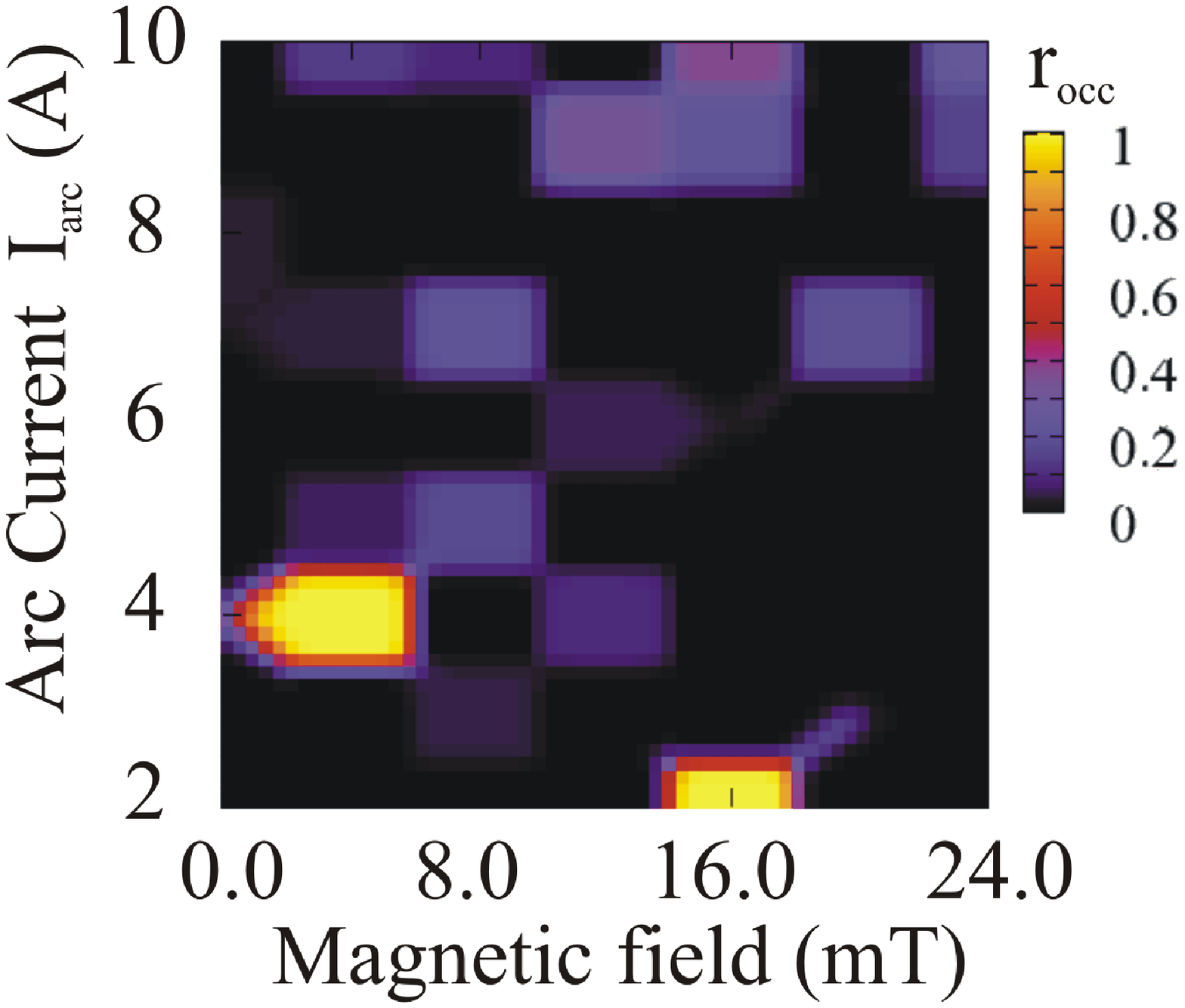}
}
\subfigure[$P=64.0~hPa$]{
\includegraphics*[width=4.2cm]{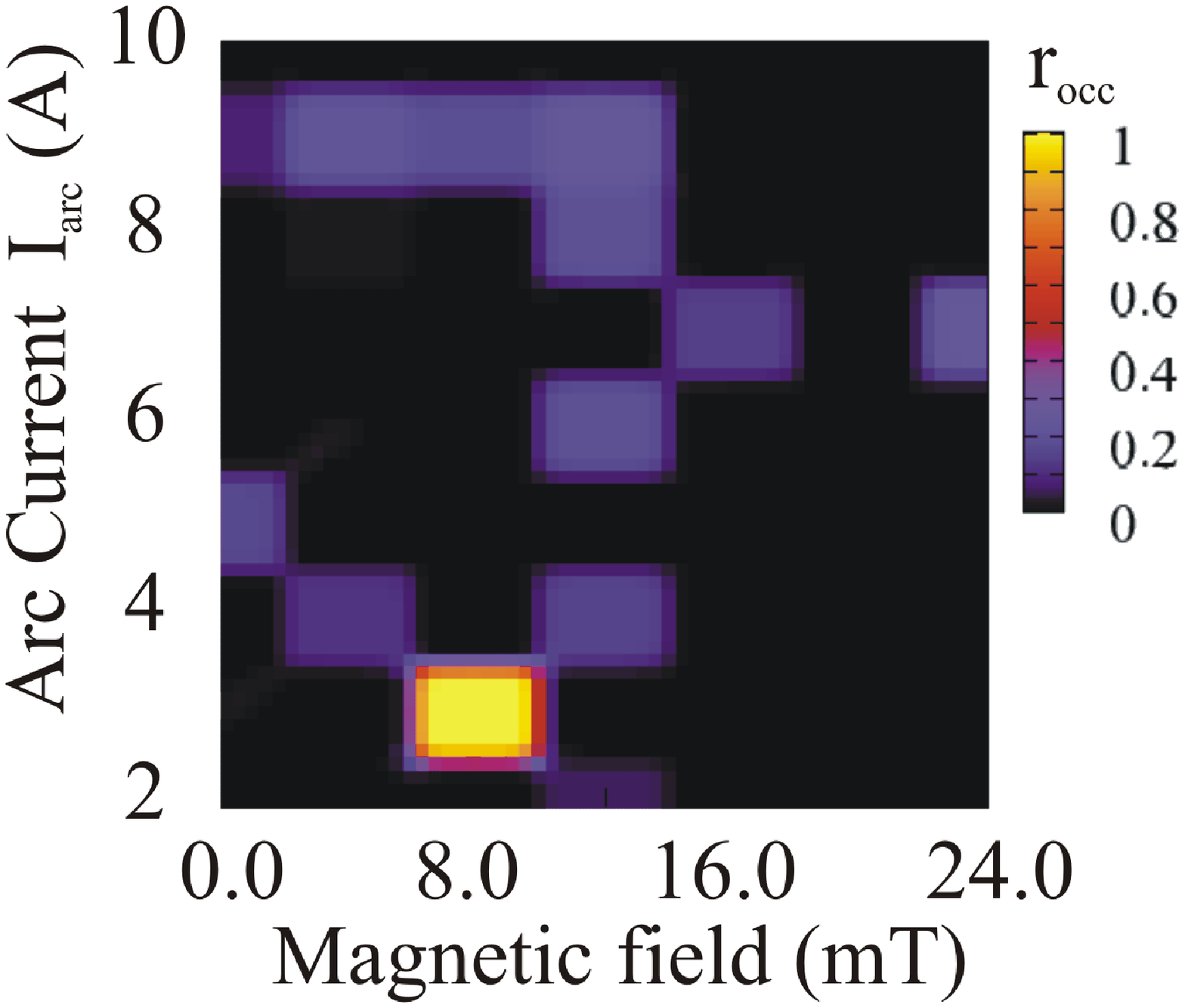}
}
\end{center}
\caption{Graph of relative occurrence of proton fraction for different gas filling pressures at constant arc potential $80~V$.}
\label{Rel_occ_vary_p}
\end{figure}

\begin{figure}[hhh]
\vspace{0.5cm}
\begin{center}
\subfigure[$V_{arc}=80~V$]{
\includegraphics*[width=4.2cm]{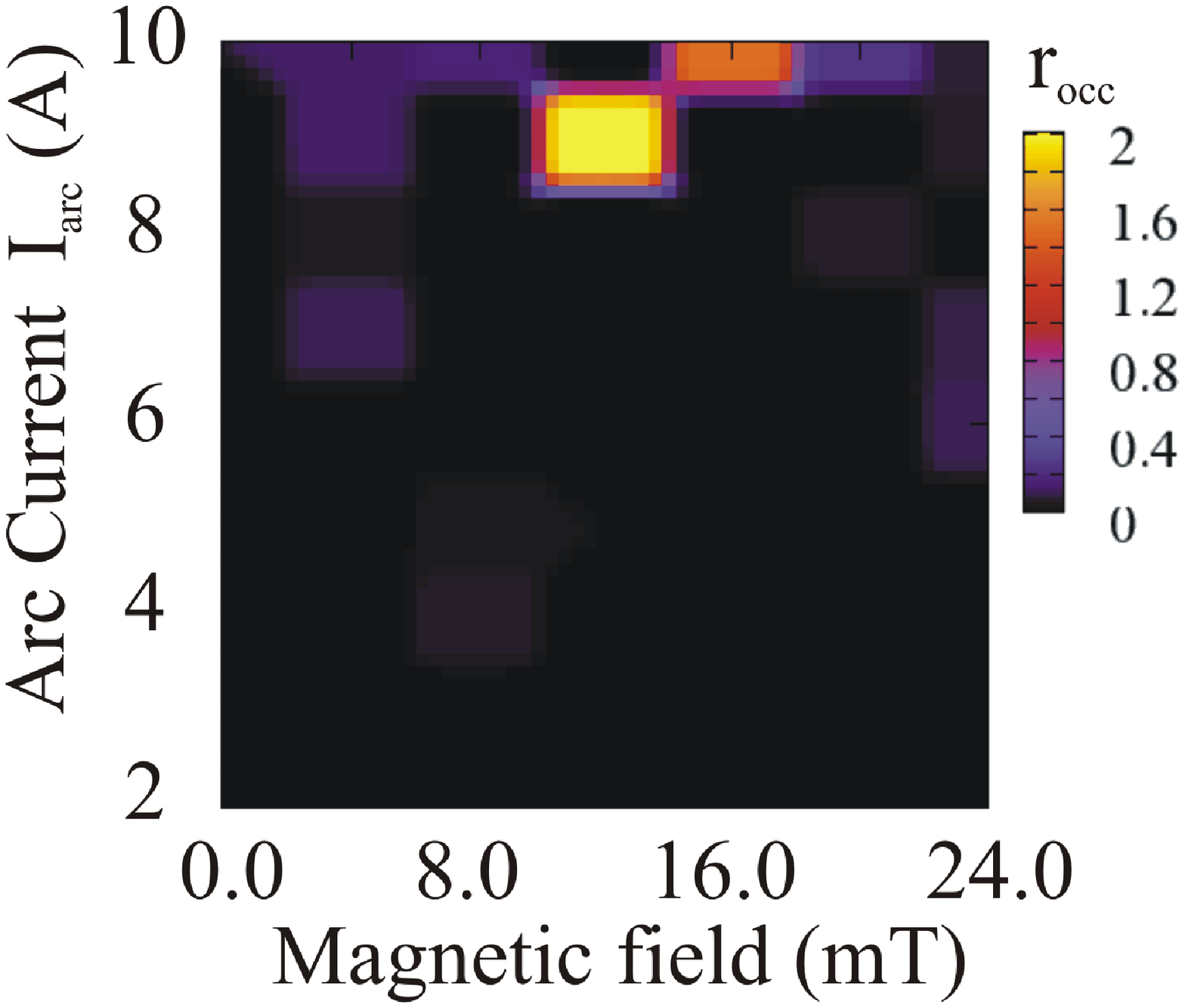}
}
\subfigure[$V_{arc}=100~V$]{
\includegraphics*[width=4.2cm]{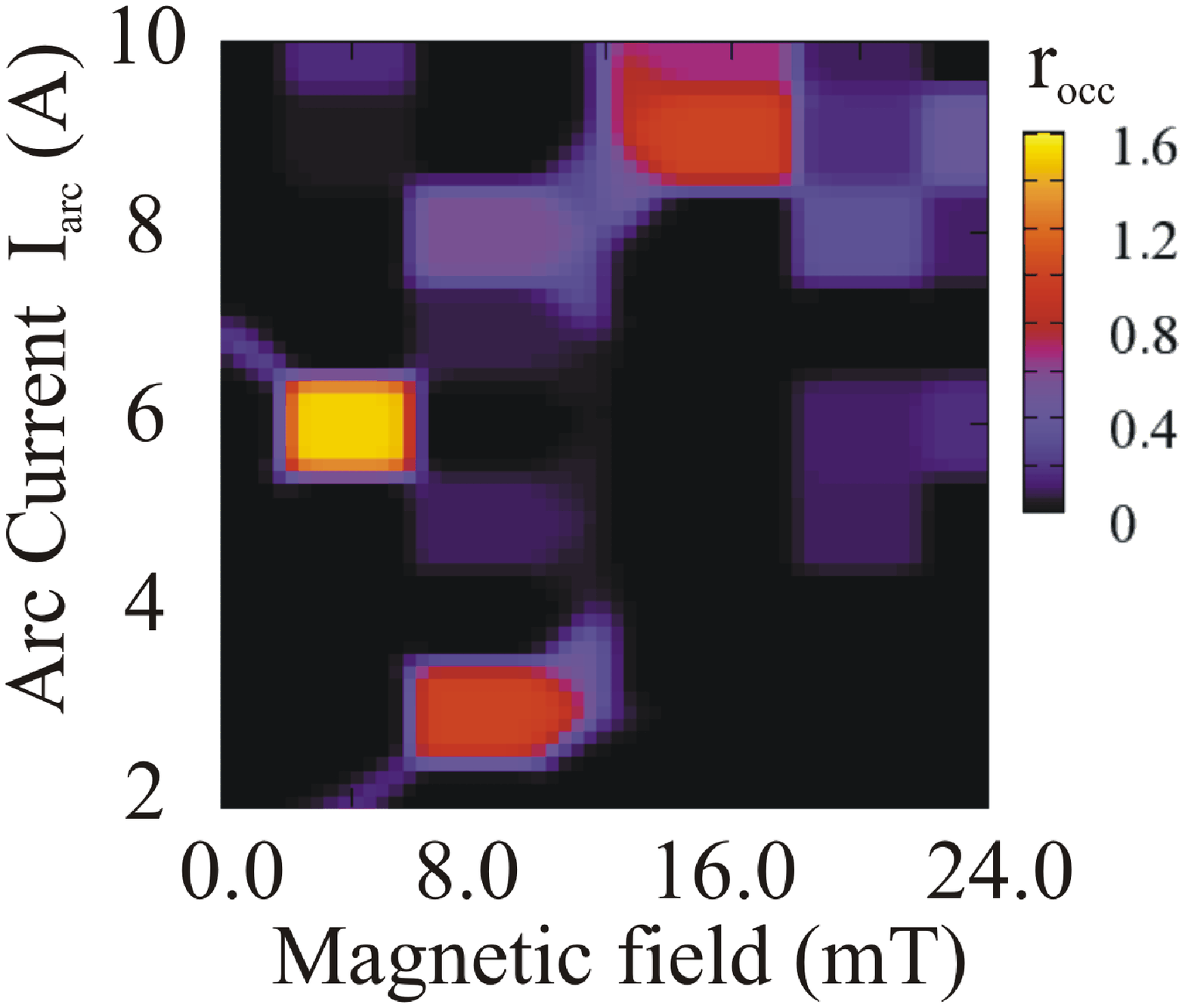}
}
\subfigure[$V_{arc}=120~V$]{
\includegraphics*[width=4.2cm]{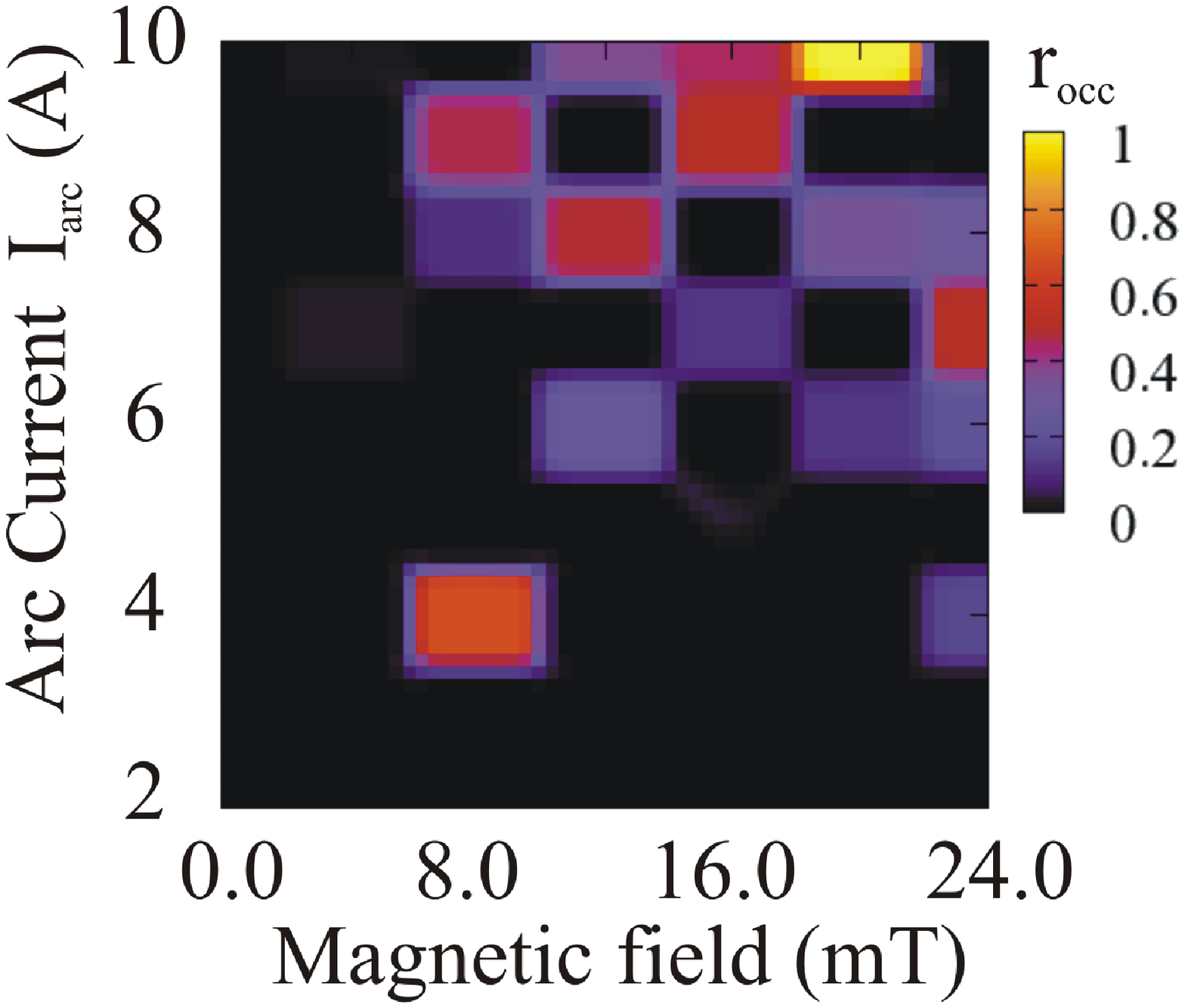}
}
\end{center}
\caption{Graph of relative occurrence of proton fraction for different arc potentials  at constant gas filling pressure $12.0~hPa$.}
\label{Rel_occ_vary_Ubogen}
\end{figure}

When these graphs were plotted for different filling pressure or arc potential the island positions moved (see Fig.\ref{Rel_occ_vary_p} and \ref{Rel_occ_vary_Ubogen}) and the intensity decreased while retaining the island structure.
% === Figure =================================================
% === Figure =================================================

% === Figure =================================================
% === Figure =================================================
Typically,  the used version of heating cathode should be changed after around $120~h$ of operation. 
\section {Conclusion}
A proton beam containing $45~to~50\%$  of $H^+$ was extracted routinely for the beam transport experiments at Frankfurt University. Especially, this ion source provides $H_{2}^+$ and $H_{3}^+$ beams with individual ratios up to $95\%$. The source can be tuned in a wide range of operating parameters to optimize the beam.

\end{document}